\documentclass[conference]{IEEEtran}
\IEEEoverridecommandlockouts
\usepackage{cite}
\usepackage{amsmath,amssymb,amsfonts}
\usepackage{algorithmic}
\usepackage{graphicx}
\usepackage{textcomp}
\usepackage{xcolor}

\usepackage{amsmath}
\usepackage{amssymb}
\usepackage{amsthm}
\usepackage{multirow}
\usepackage{booktabs}
\usepackage{subfigure}
\def\BibTeX{{\rm B\kern-.05em{\sc i\kern-.025em b}\kern-.08em
    T\kern-.1667em\lower.7ex\hbox{E}\kern-.125emX}}
\usepackage{url}
\usepackage[colorlinks, urlcolor=blue, linkcolor=blue, citecolor=blue]{hyperref}

\begin{document}

\title{Exploring Kinetic Curves Features for the Classification of Benign and Malignant Breast Lesions in DCE-MRI}

\author{\IEEEauthorblockN{Zixian Li}
\IEEEauthorblockA{\textit{SMILE Lab and MUSIC Lab} \\
\textit{School of Biomedical Engineering} \\
\textit{Shenzhen University Medical School} \\
\textit{Shenzhen University}\\
Shenzhen, 518060, China}
\and
\IEEEauthorblockN{Yuming Zhong}
\IEEEauthorblockA{\textit{SMILE Lab and MUSIC Lab} \\
\textit{School of Biomedical Engineering} \\
\textit{Shenzhen University Medical School} \\
\textit{Shenzhen University}\\
Shenzhen, 518060, China}
\and
\IEEEauthorblockN{Yi Wang$^*$}
\IEEEauthorblockA{\textit{SMILE Lab and MUSIC Lab} \\
\textit{School of Biomedical Engineering} \\
\textit{Shenzhen University Medical School} \\
\textit{Shenzhen University}\\
Shenzhen, 518060, China}
\thanks{* Corresponding author: Yi Wang, {\tt\small onewang@szu.edu.cn}.}
\thanks{
This work was supported in part by the National Natural Science Foundation of China under Grants 62071305,
in part by the Guangdong-Hong Kong Joint Funding for Technology and Innovation under Grant 2023A0505010021,
and in part by the Guangdong Basic and Applied Basic Research Foundation under Grant 2022A1515011241.
}
}

\maketitle

\begin{abstract}
Breast cancer is the most common malignant tumor among women and the second cause of cancer-related death.
Early diagnosis in clinical practice is crucial for timely treatment and prognosis.
Dynamic contrast-enhanced magnetic resonance imaging (DCE-MRI) has revealed great usability in the preoperative diagnosis and assessing therapy effects thanks to its capability to reflect the morphology and dynamic characteristics of breast lesions.
However, most existing computer-assisted diagnosis algorithms only consider conventional radiomic features when classifying benign and malignant lesions in DCE-MRI.
In this study, we propose to fully leverage the dynamic characteristics from the kinetic curves as well as the radiomic features to boost the classification accuracy of benign and malignant breast lesions.
The proposed method is a fully automated solution by directly analyzing the 3D features from the DCE-MRI.
The proposed method is evaluated on an in-house dataset including 200 DCE-MRI scans with 298 breast tumors (172 benign and 126 malignant tumors),
achieving favorable classification accuracy with an area under curve (AUC) of 0.94.
By simultaneously considering the dynamic and radiomic features, it is beneficial to effectively distinguish between benign and malignant breast lesions.
The algorithm is publicly available at~\url{https://github.com/ryandok/JPA}.
\end{abstract}

\begin{IEEEkeywords}
Breast cancer,
DCE-MRI,
Kinetic curves,
Computer assisted diagnosis 
\end{IEEEkeywords}

\section{Introduction}
Breast cancer has become the most common malignant tumor in women worldwide~\cite{ref1}, posing a great threat to female's health.
The importance of early diagnosis and treatment for breast cancer patients should be emphasized since a large proportion of patients with early breast cancer can be cured completely.
The recently widespread screening method for breast cancer is mammography, which has good resolving ability for soft tissue structure in the breast.
However, it is not adequate for most Chinese women because of their high-density breasts which reduce the sensitivity of screening~\cite{ref2}.
As an effective supplement to mammography without radiation risk, breast ultrasound (US) provides good imaging information for understanding the location, size, cystic firmness, etc.
However, the demand for experienced sonographers limits the sensitivity and specificity of screening~\cite{ref3, ref4}.

Magnetic resonance imaging (MRI) and especially the emerging methodology of dynamic contrast-enhanced (DCE)-MRI has illustrated great potential in the preoperative diagnosis and assessing therapy effects thanks to its ability to reflect the morphology, structure and dynamic characteristics of breast lesions and visualize three-dimensional (3D) high-resolution dynamic functional information, not accessible with mammography or US.
Thus DCE-MRI constitutes an important diagnostic tool for the screening of breast lesions and the identification of benign and malignant~\cite{ref25, ref5a, ref5b, ref5c, ref5d}.

\begin{figure*}[t]
	\begin{center}
		\includegraphics[width=\linewidth]{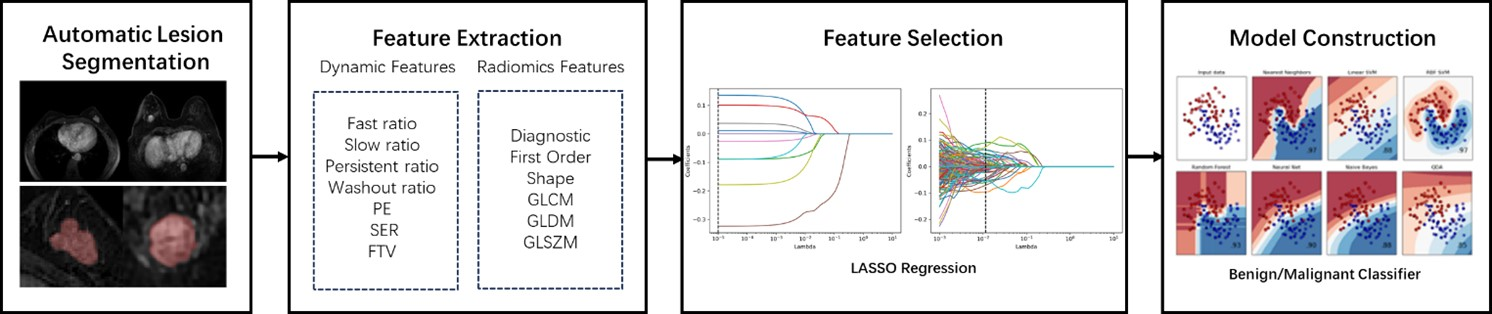}
	\end{center}
	\caption{The proposed 3D framework for the classification of benign and malignant breast tumors in DCE-MRI.}
	\label{fig:flow}
\end{figure*}

Several studies have made full use of various features in DCE-MRI scans to differentiate between benign and malignant breast tumors.
Yao~\textit{et al}.~\cite{ref6} adopted support vector machine (SVM) classifier based on texture analysis and wavelet transform on the manually selected 2D region-of-interest (ROI).
Agner~\textit{et al}.~\cite{ref7} evaluated the ability of SVM in conjunction with 2D morphological, textural, and kinetic features, yielding an accuracy of 83\% on the dataset of 41 cases.
Debbi~\textit{et al}.~\cite{ref9} employed a random forest (RF) classifier in combination with radiomic features, and achieved an accuracy of 85\%.
With the progress of convolutional neural networks (CNNs), recent studies have shifted their attention to the automatic feature-learning approach.
Rasti~\textit{et al}.~\cite{ref10} proposed a computer-assisted diagnosis system based on a mixture ensemble of 2D CNNs (ME-CNN) to distinguish between benign and malignant breast lesions, which attained an accuracy of 96\% on 112 DCE-MRI scans.
Hizukuri~\textit{et al}.~\cite{ref11} leveraged a CNN model with Bayesian optimization to analyze 2D ROI for classification and achieved an accuracy of 92\% on 56 DCE-MRI examinations.
Gravina~\textit{et al}.~\cite{ref12} introduced a 2D deforming autoencoder (DAE)-CNN to learn how to separate contrast agent effects from all the other image components when classifying breast lesions, achieving an area under curve (AUC) of 0.91 on 32 breast DCE-MRI cases.
In clinical diagnosis, radiologists usually diagnose by considering 3D information on DCE-MRI scans.
However, most studies mentioned above extracted features from 2D ROIs obtained by manual/semi-automated methods, which failed to fully utilize the rich 3D information of DCE-MRI and may even result in the absence of key details.
Therefore, more attention should be paid to the development of 3D automated algorithms for breast tumor diagnosis.

In this study, we design an automated 3D classification framework for the identification of benign and malignant breast tumors in DCE-MRI.
We leverage an automated tumor segmentation algorithm to generate the 3D ROI from the breast DCE-MRI scans.
Based on the generated tumor ROIs, we extract and analyze the 3D features representing the dynamic and radiomic characteristics to distinguish between benign and malignant tumors.
The efficacy of the proposed method is evaluated on an in-house dataset including 200 DCE-MRI scans with 298 breast tumors (172 benign and 126 malignant tumors).
Experimental results demonstrate the favorable classification accuracy by considering both 3D dynamic and radiomic features.

\section{Method}
\label{sec:method}
Fig.~\ref{fig:flow} shows the proposed automated method for the classification of benign and malignant breast tumors in DCE-MRI.
Firstly, the tumor ROIs are extracted using an automated segmentation algorithm.
Then both dynamic and radiomic features are calculated and selected for the construction of the classification model.
Finally, the constructed model can be applied for classifying benign and malignant breast tumors.

\subsection{Automated Tumor Segmentation}
\label{section:lesion segmentation}
To conduct the 3D tumor segmentation from DCE-MRI scans, we employ our previously developed joint-phase attention (JPA) network~\cite{ref15}.
The JPA network is trained using a large-scale multi-center breast DCE-MRI dataset thus is with favorable segmentation accuracy and generalization performance.
By using the JPA network, we can obtain the segmentation results of the whole breast and the tumor region (denoted as $tumor$ $mask$).

\begin{figure}[t]
	\begin{center}
		\includegraphics[width=0.95\linewidth]{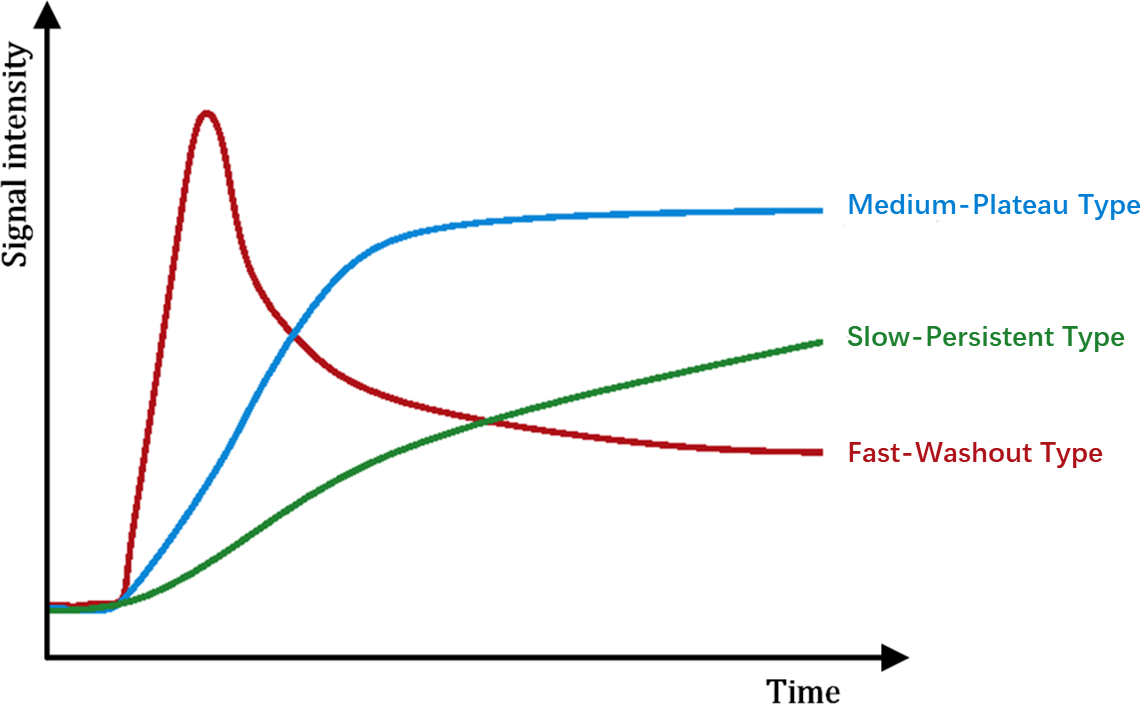}
	\end{center}
	\caption{Different types of the kinetic curves.}
	\label{fig:TIC}
\end{figure}

\subsection{Feature Extraction}
Several studies~\cite{ref18,ref19,ref20} on 2D images have proved that the dynamic features based on kinetic curves and the radiomic features based on shape and texture have strong correlation with the benign and malignant diagnosis.
We therefore consider a feature set including 3D dynamic features and radiomic features to analyze the benign and malignant breast tumors.

\subsubsection{Dynamic Features}
Dynamic features are based on the study of the permeability of diseased blood vessels by analyzing the intensity changes of lesions in T1-weighted images before and after contrast agent injection.
In the case of better vascular permeability of the lesion, contrast agent will penetrate out of the lesion, which shows decrease of the peak signal intensity in the lesion area.
Conversely, the signal intensity of the lesion area continues to increase, which will be reflected in the shape of the kinetic curve of the DCE-MRI.
The kinetic curve is the basis of DCE-MRI dynamics, which reflects the blood perfusion rate and outflow rate of breast cancer lesions.
It is divided into the initial enhancement period (with three categories based on inflow rate: slow, medium, and fast) and the delayed enhancement period (with three categories based on the outflow rate: persistent, plateau, and washout) according to the intensity reaching the peak or the time reaching 120s, as shown in Fig.~\ref{fig:TIC}.
Most benign lesions have a continuously rising shape, while malignant lesions usually have a declining shape during the delayed enhancement phase.

\begin{figure}[t]
	\begin{center}
		\includegraphics[width=\linewidth]{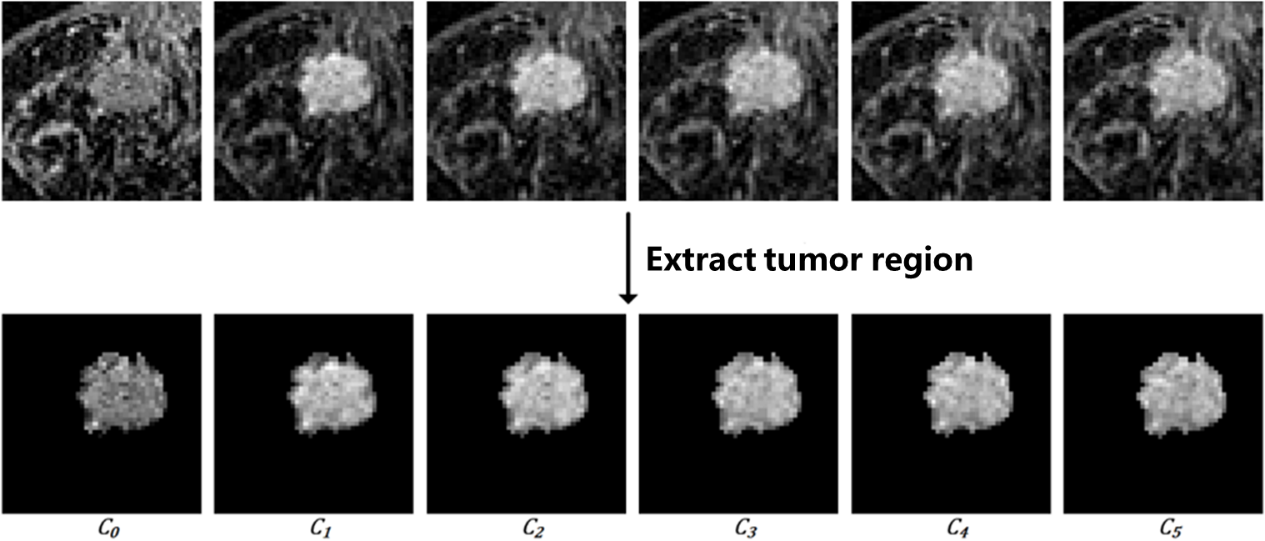}
	\end{center}
	\caption{DCE-MRI sequences and the segmented tumor regions.}
	\label{fig:sequence}
\end{figure}

Conventionally, the kinetic curve in DCE-MRI is calculated on a minimal ROI (equivalent to one dimension) which is most enhanced in the lesion.
It is generally manually selected by experienced radiologists based on the MRI signal strength.
In this study, we calculate the dynamic enhancement types of all voxels in the entire 3D tumor ROI.
Then, by analyzing the heterogeneity of the enhancement types in the lesion, each enhancement type ratio in the lesion is optimized as a dynamic feature to realize the function of benign and malignant tumor identification.
In addition, conventional dynamic features such as: the peak and average of initial peak percent enhancement (peak PE, average PE), the peak and average of signal enhancement ratio (peak SER, average SER), and functional tumor volume (FTV) are extracted as well.

Before extracting dynamic features, it is necessary to preprocess DCE-MRI data. 
Firstly, tumor ROIs in DCE-MRI sequences are extracted according to the segmentation, denoted as $C_0$, $C_1$, $\cdots$, $C_5$, as shown in Fig.~\ref{fig:sequence}.
Among them, $C_0$ is the image before contrast enhancement, $C_1$ to $C_5$ are the images from the first to the fifth phase after contrast enhancement, and the acquisition interval is 90s.
Then, the open source library Ants~\cite{ref21} is adopted for $C_0$, $C_1$, …, $C_5$ to perform fast rigid registration to correct for image misalignment due to patient movement.
According to the signal strength of $C_1$ and $C_2$, the larger one is selected as the peak enhancement period, which is recorded as $C_{peak}$.

The following is the process of calculating the heterogeneity of enhancement types in the lesion, including 3 types of initial enhancement (slow ratio, medium ratio and fast ratio) and 3 types of delay enhancement (persistent ratio, plateau ratio and washout ratio), with a total of 6 categories.
The ratio of the 3 types of initial enhancement is calculated as follows.
Firstly, the initial enhancement rate map (IERM) is calculated according to the pre-enhancement image and the peak enhancement image of the lesion, which is the voxel-to-voxel calculation:
\begin{equation}
IERM = \dfrac{C_{peak}-C_0}{C_0}\times 100\%.
\label{initia}
\end{equation}
Then according to the definition of slow, medium and fast types of initial enhancement, 3 types of masks (binary graph) are calculated:
\begin{align}
    slow\ mask &= IERM < 50\%, \\ 
    medium\ mask &= 50\% \leq IERM \leq 100\%, \\
    fast\ mask &= IERM > 100\%. 
\end{align}
Finally, the respective ratios can be calculated:
\begin{align}
    slow\ ratio &= \dfrac{\sum slow\ mask}{\sum tumor\ mask}\times100\%, \\
    medium\ ratio &= \dfrac{\sum medium\ mask}{\sum tumor\ mask}\times 100\%,\\
    fast\ ratio &= \dfrac{\sum fast\ mask}{\sum tumor\ mask}\times 100\%.
\end{align}

Regarding the calculation of delay enhancement ratios, 
the delayed enhancement rate map (DERM) is firstly calculated according to the images at the post enhancement period and the peak enhancement period:
\begin{equation}
DERM = \dfrac{C_5-C_{peak}}{C_{peak}}\times 100\%.
\label{eqn8}
\end{equation}
According to the definition of persistent, plateau and washout types of delayed enhancement, 3 types of masks are calculated:
\begin{align}
    persistent\ mask &= DERM > 10\%, \\ 
    plateau\ mask &= -10\% \leq DERM \leq 10\%, \\
    washout mask\ mask &= DERM < -10\%.
\end{align}
Then the respective ratios can be calculated:
\begin{align}
    persistent\ ratio &= \dfrac{\sum persistent\ mask}{\sum tumor\ mask}\times100\%, \\
    plateau\ ratio &= \dfrac{\sum plateau\ mask}{\sum tumor\ mask}\times 100\%,\\
    washout\ ratio &= \dfrac{\sum washout\ mask}{\sum tumor\ mask}\times 100\%. 
\end{align}

The calculation methods of the conventional dynamic features including peak PE, average PE, peak SER, average SER, and FTV are consistent with previous studies~\cite{ref21a, ref21b}.
Finally, a total of 11 digital variables including slow ratio, medium ratio, fast ratio, persistent ratio, plateau ratio, washout ratio, average PE, peak PE, average SER, peak SER and FTV are used as dynamic features to identify benign and malignant breast tumors.

\subsubsection{Radiomic Features}
\label{section:radiomics feature}
We extract 800 radiomic features composing of six categories: diagnostic features, first-order features, shape features, gray-level co-occurrence matrix (GLCM) features, gray-level size zone matrix (GLSZM) features and gray-level dependence matrix (GLDM) features~\cite{ref22}. 

In addition to extracting the radiomic features from the original image, features are extracted from the derived images after filtering and transforming.
Filters used in this study include wavelet and Laplacian of Gaussian (LoG) filters.
The wavelet filter generates eight decomposes of the original image (all combinations of high-pass (H) or low-pass (L) filters are applied in each of the three dimensions, i.e. LLL, LLH, LHL, HLL, LHH, HLH, HHL, HHH).
LoG filtering is an edge-enhanced filter that highlights areas of grayscale change.
Apart from the shape features extracted only on the original image, the rest of the radiomic features are extracted on the original image and the filtered derived images.

\begin{figure}[t]
	\begin{center}
		\includegraphics[width=\linewidth]{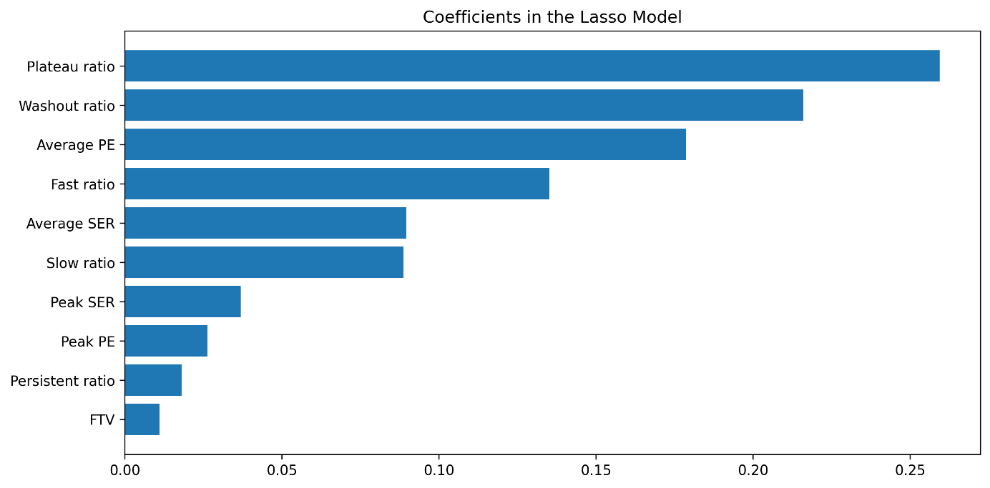}
	\end{center}
	\caption{The selected dynamic features and their corresponding coefficients on the LASSO model.}
	\label{fig:selected_D}
\end{figure}

\subsection{Feature Selection}
After extracting dynamic features and radiomic features from DCE-MRI scans, feature selection is indispensable, which enormously reduces the dimension of the feature space, screens out irrelevant features, and diminishes the training time of the classifier.
In order to achieve this goal, we optimize embedded method which combines the process of feature selection with classifier training, realizing the feature selection together with the classifier training. 
Specifically, the least absolute shrinkage and selection operator (LASSO) is employed to select dynamic features and radiomic features respectively.
At the same time, features corresponding to zero components are screened out. 
A total of 68 features composing of 10 dynamic features and 58 radiomic features are utilized to train a machine learning classifier.

\subsection{Classification}
We train the linear discriminant analysis (LDA) classifier on the combination of dynamic features and radiomic features.
LDA is a classical linear learning method, which maps samples onto the same line, making the projections of similar samples on the line as close as possible, and the projections of heterogeneous samples on the line as far as possible~\cite{ref23}.
For a given test sample, LDA maps it on the line, and finally a category prediction is made according to the position of the projection point.

\begin{figure}[t]
	\begin{center}
		\includegraphics[width=\linewidth]{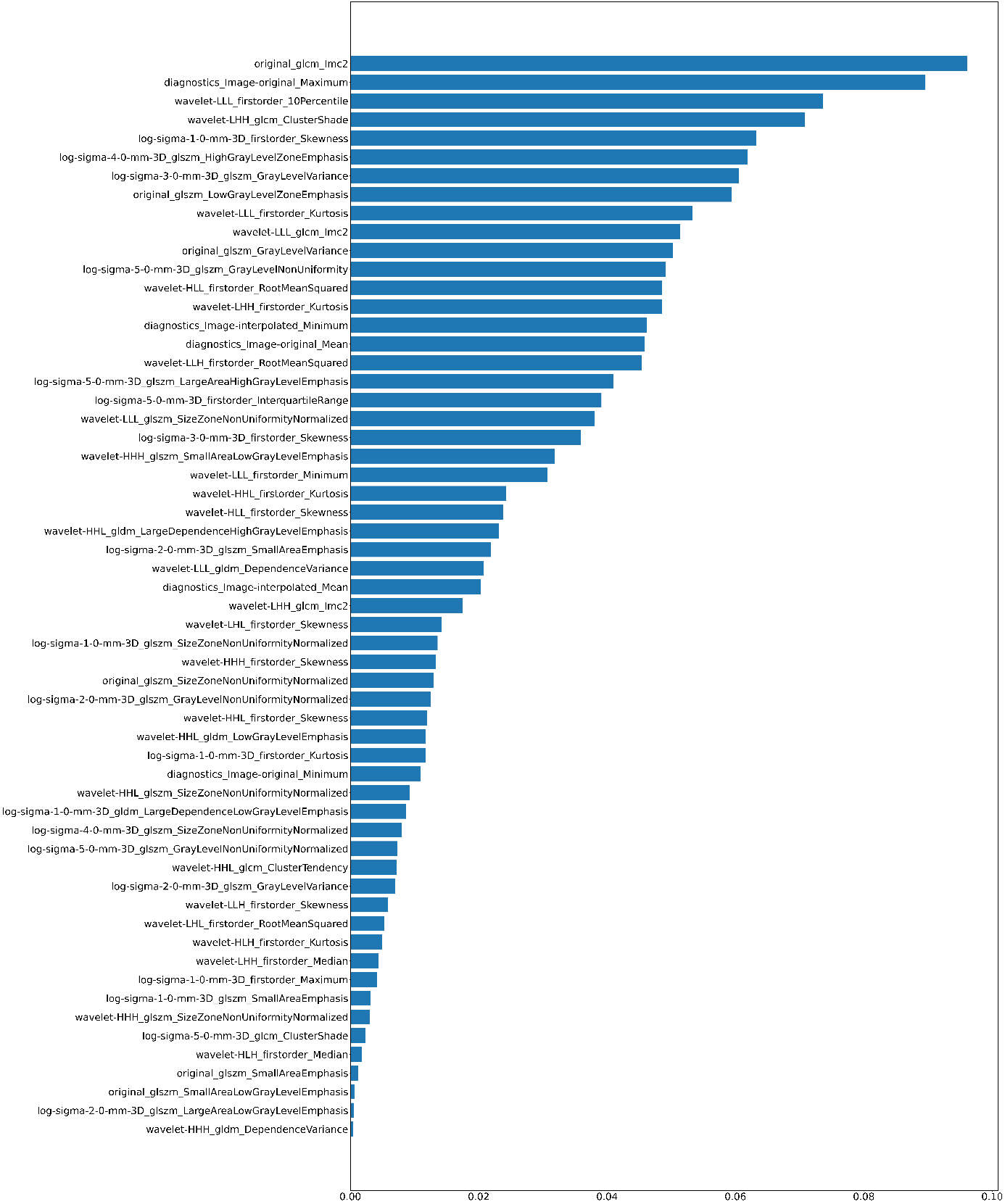}
	\end{center}
	\caption{The selected radiomic features and their corresponding coefficients on the LASSO model.}
	\label{fig:selected_R}
\end{figure}

\section{Experiments}
\label{sec:experiments}
\subsection{Materials}
This retrospective study was conducted with the approval of the Institutional Review Board of the Cancer Center of Sun Yat-Sen University, and the necessity for obtaining written informed consent was waived.
The whole collected dataset included 200 DCE-MRI scans (patients) with 298 biopsy-proven breast tumors (172 benign tumors and 126 malignant tumors).
Five-fold cross validation was conducted to investigate the classification performance.

All the breast DCE-MRI examinations were performed with a 1.5T Achieva magnetic resonance instrument (Philips) with a dedicated breast coil.
Gadolinum-based contrast agent was injected intravenously at a rate of 2ml/s at a dose of 0.2ml/kg, followed by 20ml normal saline.
DCE-MRI sequences were scanned using lipid-pressure T1-weighted gradient echo sequences (TR/TE=5.03ms/2.22ms, inversion angle=$15^{\circ}$).
The dimension of DCE-MRI volumes was $(336-432) \times (336-432) \times 300$, and the spatial resolution was $(0.926-0.949) \times (0.926-0.949) \times 0.500mm^3$.

\subsection{Evaluation Metrics}
Metrics used to quantify the classification performance included recall, precision, accuracy, F1 score, and AUC value of the area under the receiver operating characteristic (ROC) curve.
Better classification shall have larger values of these metrics.

\begin{table*}[t]
	\centering
	\caption{Classification performance of different features on LDA classifier.}
	\renewcommand{\arraystretch}{1.2}
	\label{tab:5}
	\begin{tabular}{lccccc}
		\toprule
		\textbf{Features Used} & \textbf{Accuracy} & \textbf{Recall} & \textbf{Precision} & \textbf{F1 score} & \textbf{AUC}  \\
		\midrule
		Dynamic Features & 0.7471 & 0.7583 & 0.7109 & 0.7339 & 0.8253\\
		Radiomic Features & 0.8352 & 0.8917 & 0.7810 & 0.8327 & 0.9080\\
		Combined Features & 0.8889 & 0.9417 & 0.8370 & 0.8863 & 0.9476\\
		\bottomrule
	\end{tabular}
\end{table*}

\subsection{Feature Selection Results}
\label{sec:results}
\begin{figure}[t]
	\begin{center}
		\includegraphics[width=\linewidth]{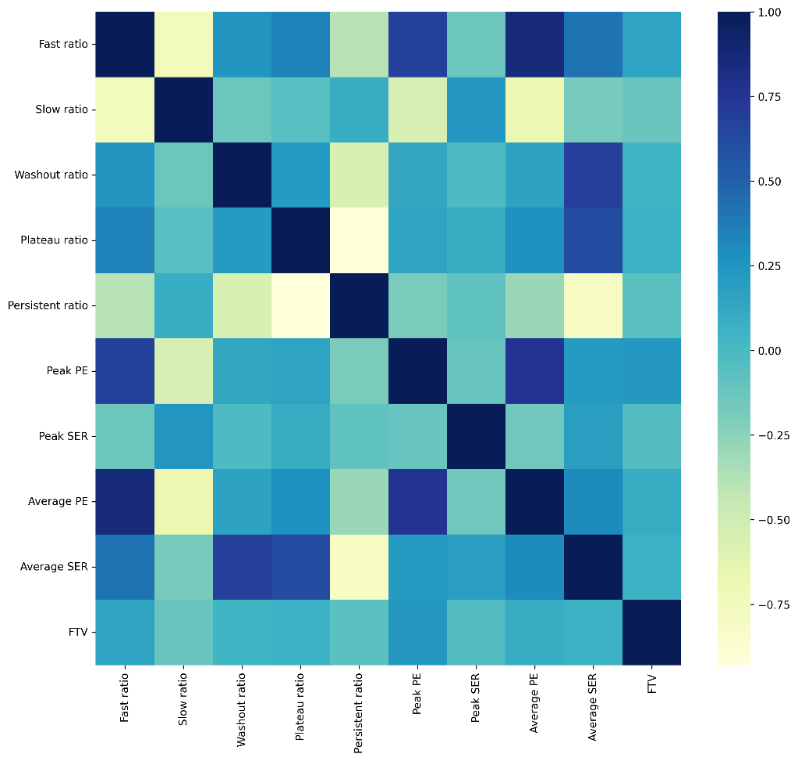}
	\end{center}
	\caption{Correlation between selected dynamic features.}
	\label{fig:correlation_D}
\end{figure}

\begin{figure}[t]
	\centering
	\includegraphics[width=\linewidth]{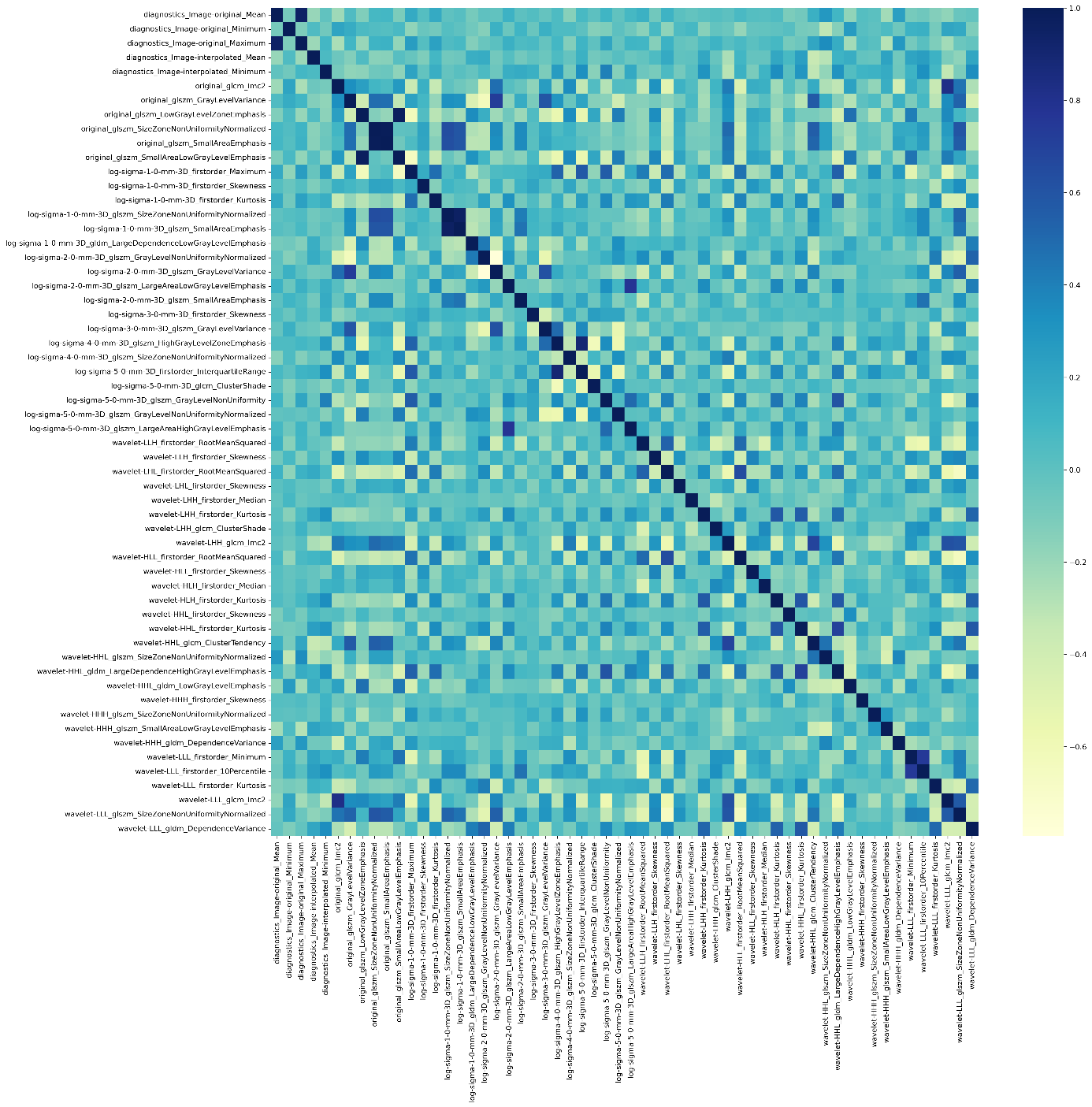}
	\caption{Correlation between selected radiomic features.}
	\label{fig:correlation_R}
\end{figure}

Fig.~\ref{fig:selected_D} and Fig.~\ref{fig:selected_R} show the features filtered by the LASSO model and their corresponding weight coefficients.
Fig.~\ref{fig:correlation_D} and Fig.~\ref{fig:correlation_R} separately show the correlation between the selected dynamic features and radiomic features, with darker colors indicating higher correlations.

A total of 10 dynamic features and 58 radiomic features were selected.
All dynamic features were selected except the medium ratio feature.
The selected radiomic features were mainly first-order features, GLCM features and GLSZM features extracted from derived images based on LoG and wavelet filtering.

\begin{figure}[t]
	\begin{center}
		\includegraphics[width=0.9\linewidth]{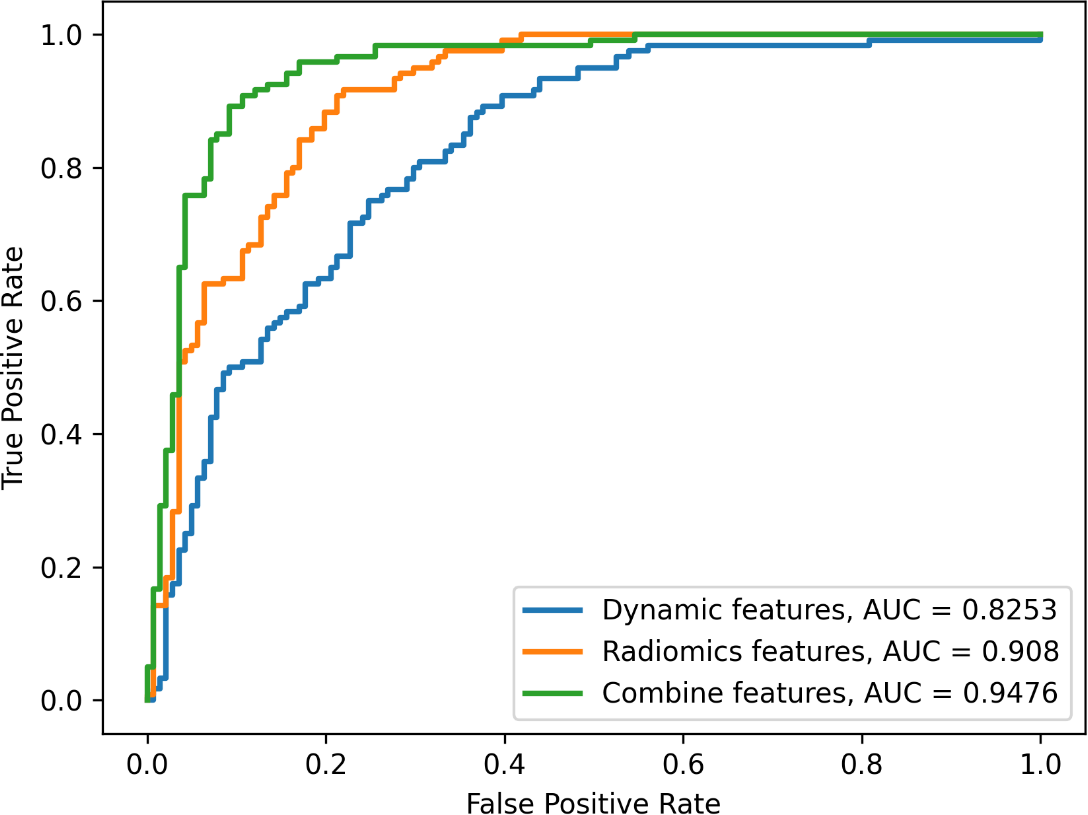}
	\end{center}
	\caption{The ROC curves of classifiers with different features.}
	\label{fig:ROC}
\end{figure}

\subsection{Classification Results}
Table~\ref{tab:5} shows the classification performance of different features on LDA classifier,
and Fig.~\ref{fig:ROC} illustrates the ROC curves.
For the dynamic feature-based classifier, only 10 dynamic features were used, and the accuracy of 0.7471, recall of 0.7583, precision of 0.7109, F1 score of 0.7339 and AUC of 0.8253 were obtained.
In contrast, a better classification performance was achieved with 0.8352 accuracy, 0.8917 recall, 0.7810 precision, 0.8327 F1 score and 0.9080 AUC by using radiomic features alone.
Finally, by combining dynamic and radiomic features, the classification performance was improved to a perceptible extent compared with only using dynamic feature or radiomic feature, yielding the best classification results with 0.8889 accuracy, 0.9417 recall, 0.8370 precision, 0.8863 F1 score and 0.9476 AUC,
which demonstrates the combined features have strong correlation with the identification of benign and malignant breast tumors.

\section{Conclusion}
\label{sec:conc}
In this study, we explore the practicality of dynamic features and radiomic features to distinguish between benign and malignant breast lesions in DCE-MRI.
To this end, we design a fully automated solution to analyze the 3D dynamic and radiomic features, and use them to construct the classification model.
The experimental results on an in-house DCE-MRI dataset show the efficacy
of the proposed method.
By simultaneously considering the dynamic and radiomic features, it is beneficial to effectively distinguish between benign and malignant breast lesions.
It is worth noting that the performance and practicability of the computer assisted diagnosis algorithms can still be enhanced.
On one hand, the 3D tumor segmentation methods can be more accurate and annotation-efficient, by further employing recently developed attention mechanisms~\cite{ref24, ref25}, or weakly-/un-supervised segmentation methods~\cite{ref26, ref27, ref28}.
On the other hand, the classification method used in this study is the simple LDA model, while future studies may consider leverage other classification algorithms~\cite{ref29}.
In addition, deep learning methods that directly analyze DCE-MRI scans and output diagnosis results can be investigated.

\end{document}